\documentclass[12pt, dvips]{article}
\def\gsim{\raise0.3ex\hbox{$>$\kern-0.75em\raise-1.1ex\hbox{$\sim$}}}
\def\lsim{\raise0.3ex\hbox{$<$\kern-0.75em\raise-1.1ex\hbox{$\sim$}}}
\begin{document}
\begin{center}
{\LARGE \bf Lorentz symmetry violation\\
\vskip 3mm
and the results of the AUGER experiment}
\end{center}
\vskip 5mm
\begin{center}
%
%
{\bf Luis Gonzalez-Mestres\\}
\vskip 3mm

{\it L.A.P.P., B.P. 110, 74941 Annecy-le-Vieux Cedex, France\\(UMR 5814 CNRS - Universit\'e de Savoie)}
\end{center}
\vskip 5mm
\begin{center}
{\large \bf Abstract\\}
\end{center}
\vspace{-0.5ex}
{\bf We briefly discuss the implications of recent AUGER results for patterns of Lorentz symmetry violation (LSV), assuming that the existence of the Greisen-Zatsepin-Kuzmin cutoff is definitely confirmed. The mass composition of the highest-energy cosmic-ray spectrum is a crucial issue. In any case, the new data allow in principle to exclude a significant range of LSV models but leave open several important possibilities : a weaker Lorentz breaking, a fundamental scale beyond the Planck scale, scenarios with threshold effects... It may even happen that spontaneous decays due to LSV fake the GZK cutoff. Space experiments appear to be needed to further test special relativity. We also comment on the consequences of AUGER data for superbradyons. If such particles are present in the Universe, they may provide new forms of dark matter and dark energy.}
\vskip 3mm
\vspace{1ex}

~

~
\section{Introduction}
\vskip 3mm \noindent
The AUGER collaboration \cite{AUG} has recently published \cite{AugerData1} and analyzed \cite{AugerData2} data on ultra-high energy cosmic rays (UHECR). The data possibly confirm the existence of the Greisen-Zatsepin-Kuzmin (GZK) cutoff \cite{GZK1,GZK2} in agreement with HiRes \cite{Hires}. But the theoretical interpretation and consequences of such a finding will also depend on other features of the experimental results. A crucial information is the composition of the highest-energy cosmic-ray spectrum. 
\vskip 3mm \noindent
It has been stated \cite{AugerData2} that cosmic rays at energies above 6 x $10^{19} ~ eV$ appear to be protons, mostly from the supergalactic plane. But this conclusion is challenged by Gorbunov et al. \cite{Gorbunov} and by Fargion \cite{Fargion}, who suggest that the most energetic observed cosmic rays are heavier objects from extragalactic sources like the nearby radio galaxy Centaurus A. More recently, the AUGER collaboration has concluded \cite{AugerData3} that the highest-energy cosmic rays are intermediate between protons and Fe nuclei, with a mean mass number of about 5 . Dermer \cite{Dermer} favors nuclei of the CNO group. Stanev \cite{Stanev} finds correlation with the updated supergalactic plane, but tends to confirm the role of Centaurus A. Obvioulsy, more statistics is needed and, eventually, separate identification of fluxes for nucleons and for different kinds of nuclei.
\vskip 3mm \noindent
The aim of this note is to briefly illustrate the relevance of this question for models of Lorentz symmetry violation that can in principle produce effects in the GZK energy region. The bounds on LSV will depend crucially on UHECR composition and on the origin of fluxes for each kind of cosmic rays. We also discuss some possible implications of the AUGER results for the superbradyon (superluminal preon \cite{gonSL}) hypothesis.
\vskip 3mm \noindent
Patterns of standard (strong) doubly special relativity \cite{SDSR} (SDSR) do not yield any testable new physics in the energy region of the predicted GZK cutoff. As, according to SDSR, the laws of Physics must be exactly the same in all inertial frames, it is possible to study the collision between a UHECR and a cosmic microwave background (CMB) photon in the rest frame of the two-particle system, where corrections to standard special relativity are too weak to play a significant role. That LSV patterns without a privileged inertial frame cannot reproduce a possible absence of the GZK cutoff was first noticed in \cite{gon02} looking in detail at the numerical predictions of the Kirzhnits-Chechin model \cite{Kir} which is of the SDSR type.
\vskip 3mm \noindent
The situation is radically different in the weak doubly special relativity (WDSR) pattern considered in our papers since 1995 \cite{gona,gonb} , where a privileged (vacuum) rest frame exists and the standard Lorentz symmetry of special relativity is assumed to be a property of the low-energy limit of a more involved scenario. The new dynamics incorporates a fundamental length scale and, possibly, a new space-time symmetry for the ultimate constituents of matter (superbradyons ?). Energy and momentum remain additive for free particles and energy-momentum conservation is preserved, but standard particles obey a deformed relativistic kinematics (DRK) that cannot be made equivalent to the conventional Lorentz symmetry. In our Universe, the laws of Physics are not identical in all inertial frames, but Lorentz symmetry remains a low-energy limit in the vacuum rest frame (VRF) for standard matter. Conventional relativity is recovered if the fundamental length is set to zero. Data from AUGER and from other UHECR experiments can test WDSR models, but not SDSR. 
\vskip 3mm \noindent
In what follows, we shall be interested in experimental tests of a simple pattern of quadratically deformed relativistic kinematics (QDRK), such as it was formulated (as a form of WDSR) in our april 1997 paper \cite{gon97a} and developed in subsequent articles \cite{gona,gonb}. This seems to be the best suited LSV model for phenomenology in the GZK region. Simple realizations of linearly deformed relativistic kinematics (LDRK, see \cite{gona,gonb}) appear to be clearly excluded by data. LDRK (QDRK) models are those where the effective parameter of Lorentz symmetry violation in the equation relating energy and momentum varies linearly (quadratically) with the particle momentum at high energy. QDRK generated at the Planck scale turns out not to be excluded by present AUGER - HiRes data, and can even produce effects (like spontaneous photon emission at ultra-high energies) able to fake the GZK cutoff.
\vskip 3mm \noindent
For simplicity, we do not consider possible thresholds \cite{gonb} in the calculations presented in this paper. If DRK incorporates thresholds at energies below Planck scale, a wider range of models will escape experimental bounds from AUGER and HiRes. We also assume that the Planck constant remains universal and is not altered by the new physics. Superbradyons, that may be the (superluminal) ultimate constituents of matter, are also considered here as possible sources of UHECR \cite{gonSL} and of large-scale cosmological phenomena.
\vskip 3mm \noindent
QDRK models were originally inspired from phonons in a lattice, where a symmetry of the Lorentz type can be generated in the low momentum limit with the speed of sound playing the role of the critical speed. The deformation of the Lorentz kinematics seems to naturally emerge in many scenarios where particles are seen as excitations of more fundamental matter (made of superbradyons ?) that interacts with a new dynamical length scale. 

\section{QDRK and the AUGER results}
\vskip 3mm 
\noindent
QDRK will be only partially excluded by the AUGER results. Many interesting possibilities will remain to be explored, as discussed below.

\subsection{The model}
\vskip 3mm 
\noindent
A simple LSV pattern with an absolute local rest frame (the VRF) for conventional matter, and a fundamental length scale $a$ where new physics is expected to originate \cite{gon97a}, is given by an equation of the form \cite{gon97a, gon97b}:

\equation
E~=~~(2\pi )^{-1}~h~c~a^{-1}~e~(k~a)
\endequation
\noindent
where $h$ is the Planck constant, $c$ the speed of light, $k$ the wave vector, and $[e~(k~a)]^2$ is a convex function of $(k~a)^2$ obtained from vacuum dynamics. Expanding equation (1) for $k~a~\ll ~1$ , we can write in the absence of other distance and energy scales \cite{gon97b}:
\begin{eqnarray}
e~(k~a) & \simeq & [(k~a)^2~-~\alpha ~(k~a)^4~
+~(2\pi ~a)^2~h^{-2}~m^2~c^2]^{1/2}
\end{eqnarray}
\noindent
$p$ being the particle momentum, $\alpha $ a model-dependent constant and {\it m} the mass of the particle. For $p~\gg ~mc$ , we get:
\begin{eqnarray}
E & \simeq & p~c~+~m^2~c^3~(2~p)^{-1}~
-~p~c~\alpha ~(k~a)^2/2~~~~~
\end{eqnarray}
\noindent
It is assumed that the Earth moves slowly with respect to the VRF and that numbers obtained in that frame for UHECR apply as well, up to very small corrections, to data on phenomena measured from Earth or from a satellite. 
\vskip 3mm \noindent
The "quadratic deformation" approximated by the expression $\Delta ~E~=~-~p~c~\alpha ~(k~a)^2/2$ in the right-hand side of (3) implies a Lorentz symmetry violation in the ratio $E~p^{-1}$ varying like $\Gamma ~(k)~\simeq ~\Gamma _0~k^2$ where $\Gamma _0~ ~=~-~\alpha ~a^2/2$ . If $c$ is a universal parameter for all particles, the QDRK defined by (1) - (3) preserves Lorentz symmetry in the limit $k~\rightarrow ~0$. 
\vskip 3mm \noindent
At energies above $E_{trans}~ \approx ~\alpha ^{-1/4} ~ (h ~ m)^{1/2} ~ c^{3/2} ~ (2 ~ \pi ~ a)^{-1/2}$, the deformation $\Delta ~E$ dominates over the mass term $m^2~c^3~(2~p)^{-1}$ in (3) and modifies all standard kinematical balances \cite{gon97c,gon97d}. Because of the negative value of $\Delta ~E$ \cite{gon97d} , it costs more and more energy, as $E$ increases, to split the incoming longitudinal momentum in the laboratory rest frame. As the ratio $m^2~c^3~(2~p~\Delta ~E)^{-1}$ varies like $\sim ~E^{-4}$ , the transition at $E_{trans}$ is very sharp. 
\vskip 3mm \noindent
Using these simple power-like laws, QDRK can lead \cite{gona, gonb} to important observable phenomena. One of them would be the suppression of the GZK cutoff. 

\subsection{The bounds}
\vskip 3mm \noindent
In terms of the equivalent fundamental energy scale $E_a ~ =~ h ~ (2 ~\pi ~a)^{-1}$ , equation (3) can be written as :
\begin{eqnarray}
E & \simeq & p~c~+~m^2~c^3~(2~p)^{-1}~
-~p~c~\alpha ~(p~ c ~E_a^{-1})^2/2~~~~~
\end{eqnarray}
\noindent
and $\Delta ~E~=~-~p~c~\alpha ~(p~ ~c ~E_a^{-1})^2/2$ , $E_{trans}~ \approx ~(m ~ E_a)^{1/2}~ c~\alpha ^{-1/4}$ . 
\vskip 5mm \noindent
Above $E_{trans}$, but also already at $E ~ \simeq ~ E_{trans}$, the standard calculations leading to the GZK cutoff do no longer hold, as the CMB photon energies used are no longer enough to produce the required reactions. The existence of the GZK cutoff, if definitely confirmed, will exclude values of $\alpha $ leading to $E_{trans}~ \lsim ~ 10^{20} ~ eV$ in the simple pattern given by (1) $-$ (3). We briefly explore here some theoretical implications of such a result.
\vskip 5mm \noindent
- {\it Protons.} For a proton, taking $E_a$ to be the Planck energy and $E_{trans}~ \approx ~ 10^{20} ~ eV$ leads to the value of $\alpha : $ $\alpha _{proton}~ \approx ~ 10^{-6}$ . This is the upper bound on the QDRK parameters that can be expected from AUGER data it the GZK cutoff is confirmed and if the UHECR turn out to be protons from nearby sources.
\vskip 3mm 
\noindent
This upper bound becomes $\alpha _{proton} ~ \lsim ~ 1$ if the fundamental energy scale lies three orders of magnitude above the Planck scale.
\vskip 5mm \noindent
- {\it Nuclei.} For a nucleus made of $N$ nucleons, the upper bound on $\alpha _{Nucleus(N)}$ becomes $\approx ~ 10^{-6}~ N^2$ using the same equations as for the proton and replacing the proton mass by that of the nucleus. For a carbon nucleus, the upper bound would be $\alpha _{carbon}~ \approx ~ 10^{-4}$ and would become $\alpha _{carbon}~ \approx ~ 1$ if $E_a$ is set to be two orders of magnitude higher than the Planck scale.
\vskip 3mm
\noindent
However, $\alpha _{carbon}~ \approx ~ 10^{-4}$ is actually the equivalent of $\alpha _{proton}~ \approx ~ 10^{-2}$ . As discussed in \cite{gon97e}, a high-energy nucleus must be regarded, to a first approximation, as a set of N nucleons where energy is additive. A simple calculation, assuming $\alpha$ to have the same value for the neutron and for the proton, leads then to $\alpha _{Nucleus(N)} ~ \approx ~ N^{-2} ~ \alpha _{proton}$. Combining this result with that obtained previously, the actual upper bound on $\alpha _{proton}$ would be $\approx ~ 10^{-6}~ N^4$ if the UHECR are nuclei with N nucleons. Therefore, if the highest-energy cosmic rays are nuclei, the bound on $\alpha _{proton}$ from AUGER data is not necessarily a very strong one and depends on the mass of the detected nuclei. The lightest component of the UHCR spectrum, if properly identified and analyzed with enough statistics, can possibly provide the most stringent bound.
\vskip 5mm 
\noindent
The composition of the UHECR spectrum is thus an essential issue to fully elucidate the theoretical implications of AUGER data, as the implications for $\alpha _{proton} $ can vary over at least four orders of magnitude. But, even assuming the UHECR are protons, the AUGER result would be far from excluding all possible versions of QDRK. Not only because there is no reason not to explore possible lower values of $\alpha $, but also for more fundamental reasons. 
\vskip 5mm 
\noindent
Actually, even admitting the validity of the concept of unification of all forces at the Planck scale, no basic Physics principle compels the physical fundamental scale to coincide with the unification scale. Exploring possible values of $E_a$ higher than the Planck energy $E_{Planck}$ is therefore not in real contradiction with the basic ideas of standard particle theory.
\vskip 5mm 
\noindent
- {\it Quarks and gluons ?} Furthermore, the question of whether $\alpha _{proton} $ is the real "basic" value of the $\alpha $ parameter must now be addressed. As hadrons are made of quarks, it may happen that, in spite of confinement, they must be dealt with as composite systems in terms quarks and gluons, similarly to nuclei in terms of nucleons. Having different values of $\alpha $ for protons and for electrons or photons could have been a source of trouble with experiment at lower energies and with higher values of $\alpha $'s, but not in the present situation. 
\vskip 3mm 
\noindent
If the basic kinematics for hadrons at ultra-high energy is that of quarks and gluons, the real QDRK of a UHECR proton or pion is likely to be a complex matter, not only because of the energies involved but also because in WDSR a $10^{20}$ $eV$ hadron is not the same physical object as the hadron at rest. Assuming confinement still occurs at ultra-high energies and taking the most conservative point of view, the value of $\alpha $ for quarks and gluons would be an order of magnitude larger than $\alpha _{proton}$. If the UHECR turn out to be carbon nuclei and similar objects, AUGER data will not be incompatible with a value of $\alpha $ for quarks and gluons, $\alpha _{QG}$, in the range $\approx ~ 0.1 ~ - ~1$ ("full-strength" LSV) taking $E_a$ to be the Planck energy. 
\vskip 5mm 
\noindent
- {\it Is the suppression of the GZK effect delayed ?} If LSV actually occurs but is weak at the Planck scale, a delayed "anti-GZK" effect is also possible. 
\vskip 5mm 
\noindent
Taking the fundamental energy to be the Planck energy, and using the bounds obtained above, one gets $E_{trans}~ \gsim ~$ 3 x $10^{21} ~ eV$ if the UHECR are protons, and $E_{trans}~ \gsim ~10^{21} ~ eV$ for carbon nuclei. As $\Delta E$ grows like $p^3$, kinematical balances can drastically change between the region 5 x $10^{19} ~ eV$ - $10^{20} ~ eV$ and energies an order of magnitude higher. There are also possible thresholds \cite{gonb} able to generate a delayed suppression of the GZK cutoff manifesting itself above the $ 10^{20} ~ eV$ region. A crucial question, in this case but also more generally, would be that of the existence of astrophysical sources that can accelerate cosmic rays to such extreme energies, or produce them otherwise. Satellite experiments seem necessary to explore the lowest possible fluxes. A complication, in this scenario, can be a possible fall of cross-sections above the GZK region, as considered in \cite{gon97f} , when $\Delta ~E$ becomes larger than the target mass.

\section{WDSR phenomenology}
\vskip 3mm 
\noindent
It seems also necessarily to briefly discuss the nature of the LSV models that UHECR experiments like Hires or AUGER can test: can WDRSR we reasonably called "doubly special relativity", even in a "weak" sense ? For this expression to be justified, we require that any observer moving at constant speed with respect to the vacuum rest frame can recognize the values of $c$ and $E_a$. Writing (1) and (2) as :
\begin{eqnarray}
E = ~ [(p ~ c)^2~-~\alpha ~(p ~ c)^4 ~ E_a^{-2} ~ +~(2\pi ~a)^2 ~ m^2~c^4]^{1/2}
\end{eqnarray}
\noindent
where, to simplify the calculation, we have also neglected higher-order terms and turned the approximation of (2) into and equality, we can write in terms of the hamiltonian $H$ the particle velocity $v$ as :
\begin{eqnarray}
v = ~ dH/dp ~ = ~ p ~ c^2 ~ E^{-1} ~ [1 ~ -~2 ~ \alpha ~(p ~ c)^2 ~ E_a^{-2}]
\end{eqnarray}
\noindent
The particle velocity in the vacuum rest frame is therefore not substantially modified, as compared to standard special relativity, if $p ~ c ~ \ll ~ E_a$ . But, as already pointed out in \cite{gon04} , the appearance of physics in the rest frame of a UHECR gets drastically changed as its energy in the VRF $E$ becomes of the same order as $E_{trans}$ or above this scale. More precisely, with suitable choices of energy and momentum units, and requiring energy and momentum to be additive and conserved in all inertial frames, we can write for $E ~ < ~ E_{trans}$ as in \cite{gon04} :
\begin{equation}
E ~ + ~ p_z ~ c ~ = ~ \Gamma ~ (E' ~ + ~ p'_z ~ c)
\end{equation}
\begin{equation}
E ~ - ~ p_z ~ c ~ = ~ \Gamma^{-1} ~ (E' ~ - ~ p'_z ~c)
\end{equation}
\noindent
where $p_z$ stands for momentum along the $z$ axis in the vacuum rest frame, $E'$ and $p'_z$ are energy and momentum along the z axis in a new inertial frame, tranverse motion is neglected and $\Gamma $ is a constant to be determined. The transition energy $E_{trans}$ is now defined as the energy for which $p_z~ = ~\alpha ^{-1/4} ~ (m ~ E_a)^{1/2}$ , so that the two last terms inside the bracket of (5) exactly cancel. With this definition, $E_{trans} = (m ~ E_a)^{1/2}~ c~\alpha ^{-1/4}$ .
\vskip 3mm
\noindent
Multiplying equations (7) and (8), we get :
\begin{equation}
(E') ^2 ~ - ~ (p'_z)^2 ~ c~^ 2 ~ = ~ (E^2 ~ - ~ p_z^2 ~ c^2) ~ = ~ m^2~c^4 ~ ~-~\alpha ~(p_z ~ c)^4 ~ E_a^{-2}
\end{equation}
If the new inertial frame is the rest frame of the particle moving with momentum $p_z$ with respect to the vacuum rest frame, equation (9) with $p'_z ~ = ~ 0$ gives the correction to the particle mass as seen in its own rest frame. There is a singularity at $E ~ = ~ E_{trans}$, where $\Gamma $ would become infinite. At $E ~ > ~ E_{trans}$ , equation (9) would suggest that the particle sees itself as a tachyon. The problem can formally be solved by a change of sign in the right-hand side of (8) above the critical energy $E ~ = ~ E_{trans}$ , but this implies a new phase of the kinematics.
\vskip 3mm
\noindent
At $E ~ < ~ E_{trans}$, the $\Gamma $ parameter can be estimated setting $p'_z ~ = ~ 0$ and using the equation :
\begin{equation}
\Gamma ^2 ~ = ~ (1 ~ + ~ p_z ~ c ~ E^{-1}) ~ (1 ~ - ~ p_z ~ c ~ E^{-1})^{-1} ~ = ~ E^{2} ~ [m^2~c^4 ~ - ~ \alpha ~(p_z ~ c)^4 ~ E_a^{-2}]^{-1}
\end{equation}
\vskip 3mm
\noindent
Eliminating $E$ and $p$ in terms of $v$, the relation between $\Gamma $ and $v$ can be obtained. But, already from (5) and (10), an important property becomes apparent. Assuming that, as suggested in previous papers \cite{gon97e} , the $\alpha $ parameter for macroscopic bodies is given by the expression :
\begin{equation}
\alpha ~ = ~ \alpha _0 ~ m_p^2 ~ M^{-2} 
\end{equation}
\vskip 3mm
\noindent
where $M$ is the mass of the macroscopic object, $m_p$ the proton mass and $\alpha _0 $ a constant of the same order as $\alpha _p $ , equation (10) becomes :
\begin{equation}
\Gamma ^2 ~ = ~ E ~ M^{-1} ~ c^{-2} ~ [m^2 ~ M^{-2} ~ - ~ \alpha _0 ~ m_p^2~ c^4 ~ E_a^{-2} ~ (p_z ~ M^{-1} ~ c^{-1})^4]^{-1}
\end{equation}
\vskip 3mm
\noindent
whereas (5) can be written as :
\begin{equation}
E ~ M^{-1} ~ c^{-2} ~ = ~ [(p_z ~ M^{-1} ~ c^{-1})^2 ~ - ~ \alpha _0 ~ m_p^2~ c^4 ~ E_a^{-2} ~ (p_z ~ M^{-1} ~ c^{-1})^4 ~ +~(2\pi ~a)^2]^{1/2}
\end{equation}
\vskip 3mm
\noindent
It therefore follows that in this limit, and with the above assumption, (5) becomes an equation relating the two dimensionless variables $\Sigma ~ = ~ E ~ M^{-1} ~ c^{-2}$ (reduced energy) and $\Pi ~ = ~ p_z ~ M^{-1} ~ c^{-1}$ (reduced momentum), and that $\Gamma $ is a function of these reduced variables for a deformed boost from the vacuum rest frame to the inertial frame defined by the free motion of the macroscopic object. The same result is obtained for $v$ , using (6). One can therefore write for conventional macroscopic bodies : 
\begin{equation}
v ~ = ~ v ~ (\Sigma ~ , ~ \Pi) ~ ~  ; ~ ~ \Gamma ~ = ~ \Gamma ~ (\Sigma ~ , ~ \Pi)
\end{equation}
\vskip 3mm
\noindent
The relation between $\Gamma $ and $v$ will therefore not depend on $M$ and, just as in standard special relativity, the deformed Lorentz transformation can be given a universal definition independent of the mass of the macroscopic body considered. Thus, the existence of a transition energy $E_{trans} $ amounts, for macroscopic objects, to a universal transition speed in vacuum $v_{trans}$ that can be determined from the above equations. Similarly, it can be checked that macroscopic observers with rest inertial frames moving with respect to the vacuum rest frame at $v ~ < ~ v_{trans}$ and at speed low enough for conventional bodies to exist, are in principle able to uniquely determine the values of $c$ and of the dimensionless parameter $\alpha _{macro} ~ = ~ \alpha _0 ~ m_p^2~ c^4 ~ E_a^{-2}$ provided they are able to perform precise enough experiments. However, as $m_p^2~ c^4 ~ E_a^{-2} ~ \approx ~ 10^{-38}$ if $ E_a$ is close to Planck scale, WDSR effects seem very difficult to observe for macroscopic objects.
\vskip 3mm 
\noindent
For the elementary particles of standard theory, it seems impossible to set a universal law $\alpha ~ \propto ~ m^{-2}$ , just because of the existence of massless and nearly massless particles. No equivalent of (14) exists at such scales where the formal definition of a boost from the vacuum rest frame to a moving frame will depend on the nature of the object carrying the inertial reference system. Therefore, WDSR is a "weak" form of special relativity in the following sense: $\it i)$ it incorporates a vacuum rest frame extremely difficult to identify experimentally, two fundamental constants (the speed of light $c$ and the fundamental energy scale $E_a$) and a dynamical function $f ~ (p ~ c ~ E_a^{-1})$ describing the breaking of Lorentz symmetry at energies scales below $E_a$ ; $\it ii)$ it allows for a "geometric" formulation only in the limit of macroscopic bodies.

\section{QDRK, the composite proton and nuclei}
\vskip 3mm 
\noindent
The aim of this Section is to further discuss the point about quarks and gluons evoked in Section 2 . As emphasized in previous papers (e.g. \cite{gon97d}), all stable elementary particles must in principle have the some value of $\alpha $ up to small corrections. Otherwise, those with lower $\alpha $ would undergo spontaneous decays or spontaneously radiate in vacuum. This requirement becomes less stringent if significant bounds can be set on the LSV parameters.
\vskip 3mm 
\noindent
Assume, for instance, that $\alpha _{proton} ~ = ~ \Lambda ~ \alpha _0 $ , with $0 ~ \le ~ \Lambda ~ < ~1 $ , $\alpha _0 $ being the value of $\alpha $ for the photon and leptons. In a calculation with one space dimension, a proton with momentum $p$ can spontaneously radiate a photon with momentum $p_1$ if :
\begin{equation}
\alpha _0 ~ p_1^3 ~ (c ~ E_a^{-1})^2 ~> ~ (m ~ c)^2 ~ [(p ~ - ~ p_1)^{-1} - p^{-1}] ~ + ~ \Lambda ~ \alpha _0 ~ [p^3 ~ - ~ (p ~ - ~ p_1)^3] ~ (c ~E_a^{-1})^2
\end{equation}
$m$ being the proton mass. Then, an absolute condition for spontaneous decay to be kinematically allowed at some energy is :
\begin{equation}
G ~(\zeta) ~ = ~ \zeta ^3 ~ - ~ \Lambda ~ [1 ~ - ~ (1 ~ - ~ \zeta ) ^3] ~> ~ 0
\end{equation}
where $\zeta  ~ = ~ p_1 ~ p^{-1}$, and $\zeta $ must be in the range $0 ~ < ~ \zeta ~ < ~ 1$ . $G ~(\zeta)$ is negative for small $\zeta $ but equals $1 ~ - ~ \Lambda $ at $\zeta ~ = ~ 1$ . It changes sign at $\zeta _0 ~ = ~ [(12 ~ \Lambda ~ - 3 ~ \Lambda ^2)^{1/2} ~ + ~ 3 ~ \Lambda ] ~ [2 ~ (1 ~ - ~ \Lambda )]^{-1}$. 
Thus, the allowed range of values of $\zeta ~ $ for spontaneous photon emission is very unconventional and independent of the energy of the particle.
\vskip 3mm
\noindent
The spontaneous decay is kinematically allowed for $\zeta ~ > ~ \zeta _0 $ and :
\begin{equation}
\alpha _0 ~ p{^4} ~ (m ~E_a)^{-2} ~> ~ \zeta ~ [(1 ~ - \zeta ) ~ G ~(\zeta )] ^{-1} ~ = ~ [(1 ~ - \zeta ) ~ (\zeta ^2 ~ - \zeta ^2 ~ \Lambda ~ + 3 ~ \zeta ~ \Lambda ~ - ~ 3 ~ \Lambda )]^{-1}
\end{equation}
\vskip 3mm
\noindent
The numerical values of $\zeta _0 $ are : $\simeq ~ 0.75 $ for $\Lambda ~ \sim ~ 0.1$ ; $\simeq ~ 0.19 $ for $\Lambda ~ \sim ~ 0.01$ ; $\simeq ~ 0.055 $ for $\Lambda ~ \sim ~ 10^{-3}$ ... In all cases, the energy of the emitted photon will be atypically large as compared with the quark energies involved and with standard emission of photons by nuclei. The process has no equivalent in special relativity, and a particle at rest in the VRF cannot be used as a reference. Therefore, we may expect unusually small matrix elements for this new kind of decays. This must be taken into account for phenomenological applications. 
\vskip 3mm
\noindent
Kinematically, assuming $\alpha _0 $ to be in the range $0.1 - 1$ for quarks, leptons and gauge bosons, such decays would be allowed for protons or nuclei above some scale between $E ~ \sim ~ 10^{19} ~ eV$ (where the Lorentz factor is $\sim ~ 10^{10} $ for protons) and the GZK energy. It may happen, however, that, because of matrix elements and phase space, they manifest themselves only at higher energies and/or at very large distance scales. 
\vskip 3mm
\noindent
It is therefore not impossible that spontaneous decays of UHECR (protons and nuclei) by this new kind of photon emission fake the GZK cutoff at very large distance scales in the Universe without changing the situation at the $\sim ~ 100 ~ Mpc $ scale. In this case, the predicted GZK cutoff would actually be suppressed, but Lorentz symmetry violation would produce a similar effect by other means. This question obviously deserves further investigation.
\vskip 3mm
\noindent
The scenario considered here is different from that studied in \cite{gon001} where we discussed the inhibition of synchrotron radiation in LSV patterns, leading to possible tests like that presented later by other authors \cite{Jacobson} for a version of LDRK. In the present case, charged particles accelerated to ultra-high energies may become able to emit photons in the energy range considered here which is not the same as that of synchrotron radiation. Similar to the previous discussion, we expect UHECR acceleration to be much faster than the lifetime for spontaneous decay.

\section{Superbradyons}
\vskip 3mm 
\noindent
Superbradyons \cite{gonSL} are particles with positive mass and energy, and a critical speed in vacuum much larger than the speed of light. A simple assumption is that they have their own internal symmetries and obey a new Lorentz symmetry, with the new critical speed $c_s$ replacing $c$. 
\vskip 3mm 
\noindent
Such superluminal particles can be the ultimate constituents of matter and of the physical vacuum in a similar way to preon models but feeling different properties of space and time, just as a photon or a free nucleon do not see the same space and time structure as a phonon in a solid. No basic physics principle can {\it a priori} compel the constituents of matter at a given scale to feel the same space-time properties (for instance, through the relation between energy and momentum for a free object) as the excitations of the composite matter.
\vskip 3mm 
\noindent
Thus, our standard particles would look to some extent, in a universe made of superbradyons, similar to phonons and quasiparticles in a body made of conventional matter. They would be excitations of a ground state (the physical vacuum) made of the new fundamental matter. Superbradyons can possibly produce directly several kinds of observable effects, if they exist inside our standard Universe as individual constituents of the vacuum or as free particles. If they can exist as free particles, the new fundamental superluminal constituents would always have values of $E~p^{-1}$ much larger than $c$ . They would be in principle able to emit "Cherenkov" radiation in the form of very high-energy standard particles. 
\vskip 3mm 
\noindent
It may also happen that free superbradyons present in our Universe have already lost all their energy available for "Cherenkov" radiation and form a dark matter sea. More precisely, a "relativistic" superbradyon ($v ~ \sim ~ c_s$) would be kinematically allowed to emit conventional particles in vacuum with a characteristic signature related to the property $E~p^{-1}~\gg ~ c$ (back to back or isotropic events). But a simple calculation shows that "nonrelativistic" superbradyons ($v ~\ll ~ c_s$) can also emit "Cherenkov" radiation in vacuum if their speed is larger than $c$. Writing, for a "nonrelativistic" superbradyon :
\vskip 3mm
\noindent
\begin{equation}
p ~ \simeq ~ m ~ v
\end{equation}
\begin{equation}
E ~ \simeq ~ m ~ c_s ^2 ~ ~ + ~ m ~ v^2/2
\end{equation}
\vskip 3mm
\noindent
the energy loss for longitudinally emitting a massless bradyon with momentum $p'$ is less than $p' ~ c$ for $v ~ < ~ c$ . Below this speed, spontaneous decays can no longer preserve energy and momentum conservation, and the superbradyon ceases to emit conventional particles. It is therefore possible that our Universe contains a sea of superbradyons with speed close to $c$ , if they can exist as free particles. In all cases, a nontrivial question will be that of the interaction of such superbradyons with the physical (superbradyonic) vacuum in our Universe. 
\vskip 3mm
\noindent
If superbradyons have not yet lost all their available energy for spontaneous radiation, they can possibly be detected by experiments like AUGER \cite{gonSL} . A clear identification would in principle be difficult for cosmic rays from superbradyonic decays originated far from our atmosphere, but a superbradyonic interaction in the atmosphere would have a clear signature. Satellite experiments should in principle be more sensitive to possible (rare) events from superbradyons. But it may happen that, for basic physical reasons, superbradyons are present near standard cosmic accelerators and radiate from the same places where conventional acceleration occurs.
\vskip 3mm
\noindent
Superbradyons open the way to a new cosmology, with potentially a new approach to inflation and to the cosmological constant problem as well as new forms of dark matter and dark energy. Similarly, our physical vacuum can actually be full of superbradyons forming a new kind of condensate instead of the conventional Higgs particles. A crucial question would be in both cases that of the gravitational properties of superbradyons, not only as free particles in our vacuum but also as constituents of the "Higgs" condensate.
\vskip 3mm
\noindent
It was suggested in our 1995 papers \cite{gon95} that superbradyons may have produced in the early universe effects similar to those attributed to inflation, and that their cosmological role could replace the standard inflationary pattern. As light is $10^6$ times faster than sound, nothing prevents in principle the critical speed of free superbradyons from being, for instance, $10^{20}$ times larger than $c$ . Combining this property with "Cherenkov" emission for $v ~ > ~ c$ and with the fact that superbradyons will not be coupled to gravitation in the same way as "normal" matter, a consistent pattern avoiding also the cosmological constant problem can hopefully emerge. The possibility that superbradyons are dark matter was already considered in \cite{SLDM} . The same approach, together with the superbradyonic nature of the physical vacuum, should be able to solve the dark energy question. More details on this point will be given elsewhere.
\vskip 3mm
\noindent
As already emphasized in our previous papers, superbradyons are fundamentally different from tachyons, as the later preserve standard Lorentz invariance whereas superbradyons explicitly violate it and have a different critical speed in vacuum. Superbradyon cosmology will therefore not involve the same basic mechanisms as the models using tachyons proposed since 2001 \cite{TachyonsCosmo}. It must also be noticed that, already in standard physics, the possible role of curvature and matter inhomogeneities in our Universe is not fully understood \cite{Buchert} . If superbradyons exist, their distribution and their collective interaction properties will also be major issues.

\section{Conclusion and comments}
\vskip 3mm 
\noindent
Commenting on the relativity principle he had formulated as early as 1895  \cite{Poincarea}, the French mathematician Henri Poincar\'e wrote in 1901 \cite{Poincareb} : 
\vskip 2mm
{\it "This principle will be confirmed with increasing precision, as measurements become more and more accurate"}.
\vskip 3mm
\noindent
Twenty years later, Albert Einstein wrote in "Geometry and Experiment" \cite{Einstein}:
\vskip 2mm
{\it "The interpretation of geometry advocated here cannot be directly applied to submolecular spaces... it might turn out that such an extrapolation is just as incorrect as an extension of the concept of temperature to particles of a solid of molecular dimensions"}. 
\vskip 3mm
\noindent
To date, Poincar\'e's prediction remains experimentally valid. For reasons that are far from obvious, the sensible word of caution by Einstein has not yet been confirmed by data at the highest available energies. This question is, today, the most fundamental problem of Particle Physics and Astrophysics. To fully study it, space experiments are required to complete the AUGER search. Cosmological implications need also to be explored.
\vskip 5mm 
\noindent
The essential issue of the composition of the UHECR cosmic-ray spectrum seems to need further clarification with better statistics in current experiments. Whatever the result, AUGER and Hires data are ruling out a significant range of models and values of the basic parameters of possible Lorentz symmetry violation patterns. But another important range will remain to be tested experimentally. Quadratically deformed relativistic kinematics provides an example of models that AUGER and Hires results will not allow to exclude, even if a range of parameters will indeed become inconsistent with experiment. 
\vskip 5mm 
\noindent
As explained above, if quarks and gluons are the relevant elementary particles to build the QDRK for hadrons and the UHECR are light nuclei, the strength of LSV can even be close to 1 at the Planck scale without conflicting with AUGER data. LSV is even able to generate effects similar to the GZK cutoff but with a completely different dynamical origin. In any case, it clearly follows from the above analysis that the experimental exploration of possible Lorentz symmetry violations at ultra-high energy must necessarily be pursued beyond AUGER, in order to further explore Physics as close as possible to the Planck scale.
\vskip 5mm
\noindent
Rather than closing a debate, AUGER results are likely to open new ways for research in the field. Not only within existing phenomenological patterns of LSV, but also beyond these conservative approaches. Two classes of LSV scenarios can be considered : 
\vskip 5mm
\noindent
- {\it Lorentz symmetry violation within present standard particle physics}. LSV occurs in the framework of present particle physics theory, without fundamentally changing it. It may originate from quantum gravity or from some other domain of the dynamics, at the Planck scale or at some fundamental scale not far from the Planck scale or beyond it. HiRes and AUGER data will play an important role in constraining the nature and the parameters of many models, but other possibilities will still be left open.
\vskip 5mm
\noindent
- {\it Fundamentally new physics}. Standard particle physics and cosmology is not necessarily the ultimate theory, especially if the real fundamental scale lies beyond Planck scale. The particles of the present standard theory are not necessarily the elementary constituents of matter, and $c$ may not be the ultimate critical speed in vacuum. Our Universe may for instance be a superconducting bubble in superluminal preon (superbradyonic) matter \cite{gonSL}, and the present vacuum, a ground state inside the bubble. Then, $a$ would be in a sense similar to a sort of lattice step or correlation length of a "quasiparticle" and "phonon" model in this superbradyonic matter \cite{gon97a}. Worries about the cosmological constant and dark energy would hopefully be solved in this way, taking into account the properties of the new fundamental constituents of matter. Superbradyons may even exist as free particles in our universe, possible travelling at a speed close to $c$ after having released all the kinematically allowed radiation.
\vskip 5mm
\noindent
Obviously, none of the patterns just mentioned is free from speculation. But concluding, at the present stage, that nothing unconventional happens beyond the reach of present data would also involve a substantial amount of speculation. Globally, many fundamental pieces of information are still missing. Therefore, further experiments are required in order to complete the work of the HiRes and AUGER collaborations and eventually open the way to a full understanding of physics at the Planck scale or, if needed, beyond it.

\end{document}